\documentclass[fleqn,usenatbib]{mnras}

\usepackage{newtxtext,newtxmath}

\usepackage[T1]{fontenc}

\DeclareRobustCommand{\VAN}[3]{#2}
\let\VANthebibliography\thebibliography
\def\thebibliography{\DeclareRobustCommand{\VAN}[3]{##3}\VANthebibliography}

\usepackage{graphicx}	
\usepackage{amsmath}	
\usepackage{soul}
\usepackage{orcidlink}
\usepackage{booktabs}
\usepackage{physics}
\usepackage{comment}
\usepackage{soul}

\usepackage{acronym}

\newacro{ADM}[ADM]{Arnowitt–Deser–Misner}
\newacro{ASD}[ASD]{amplitude spectral density}
\newacro{ANN}[ANN]{artificial neural network}
\newacro{CNN}[CNN]{convolutional neural network}
\newacro{aLIGO}[aLIGO]{Advanced LIGO}
\newacro{BBH}[BBH]{binary black hole}
\acrodefplural{BBH}[BBHs]{binary black holes}
\newacro{BH}[BH]{black hole}
\acrodefplural{BH}[BHs]{black holes}
\newacro{BNS}[BNS]{binary neutron star}
\acrodefplural{BNS}[BNS]{binary neutron stars}
\newacro{BSSN}[BSSN]{Baumgarte-Shapiro-Shibata-Nakamura}
\newacro{CCSN}[CCSN]{core-collapse supernova}
\acrodefplural{CCSN}[CCSNe]{core-collapse supernovae}
\newacro{CBC}[CBC]{compact binary coalescence}
\acrodefplural{CBC}[CBCs]{compact binary coalescences}
\newacro{CE}[CE]{Cosmic Explorer}
\newacro{CFL}[CFL]{Courant-Friedrichs-Lewy}
\newacro{CFS}[CFS]{Chandrasekhar-Friedman-Schutz}
\newacro{CMB}[CMB]{Cosmic Microwave Background}
\newacro{CI}[CI]{credible interval}
\acrodefplural{CI}[CIs]{credible intervals}
\newacro{cWB}[cWB]{Coherent Wave Burst}
\newacro{DL}[DL]{deep learning}
\newacro{DNS}[DNS]{direct numerical simulations}
\newacro{EM}[EM]{electromagnetic}
\newacro{ETG}[ETG]{event trigger generator}
\newacro{ET}[ET]{Einstein Telescope}
\newacro{EOB}[EOB]{effective one-body}
\newacro{EOS}[EOS]{equation of state}
\acrodefplural{EOS}[EOSs]{equations of state}
\newacro{FAP}[FAP]{false alarm probability}
\newacro{FAR}[FAR]{false alarm rate}
\newacro{FFE}[FFE]{force-free electrodynamics}
\newacro{FFT}[FFT]{fast Fourier transform}
\newacro{GR}[GR]{General Relativity}
\newacro{GRMHD}[GRMHD]{general-relativistic magnetohydrodynamics}
\newacro{GRB}[GRB]{gamma-ray burst}
\acrodefplural{GRB}[GRBs]{gamma-ray bursts}
\newacro{GWTC3}[GWTC-3]{}
\newacro{GWTC}[GWTC]{Gravitational-wave Transient Catalog}
\newacro{GW}[GW]{gravitational wave}
\acrodefplural{GW}[GWs]{gravitational waves}
\newacro{HLL}[HLL]{Harten-Lax-van Leer}
\newacro{HLLE}[HLLE]{Harten, Lax, Van Leer, Einfeldt}
\newacro{HMNS}[HMNS]{hypermassive neutron star}
\acrodefplural{HMNS}[HMNSs]{hypermassive neutron stars}
\newacro{KH}[KH]{Kelvin -- Helmholtz}
\newacro{KHI}[KHI]{Kelvin -- Helmholtz instability}
\newacro{LES}[LES]{large-eddy simulations}
\newacro{LIGO}[LIGO]{Laser Interferometer Gravitational-Wave Observatory}
\newacro{LVK}[LVK]{LIGO, Virgo, and KAGRA}
\newacro{MRI}[MRI]{magnetorotational instability}
\newacro{MHD}[MHD]{magnetohydrodynamic}
\newacro{MInIT}[MInIT]{MHD-Instability-Induced-Turbulence}
\newacro{MMA}[MMA]{multi-messenger astronomy}
\newacro{MP}[MP]{monotonicity-preserving}
\newacro{NR}[NR]{Numerical Relativity}
\newacro{NS}[NS]{neutron star}
\acrodefplural{NS}[NSs]{neutron stars}
\newacro{NSBH}[NSBH]{neutron star - black hole}
\newacro{O2}[O2]{second observing run}
\newacro{O3}[O3]{third observing run}
\newacro{O4}[O4]{fourth observing run}
\newacro{O5}[O5]{fifth observing run}
\newacro{PI}[PI]{parasitic instability}
\acrodefplural{PI}[PIs]{parasitic instabilities}
\newacro{PPM}[PPM]{Piecewise Parabolic Method}
\newacro{PNS}[PNS]{protoneutron star}
\acrodefplural{PNS}[PNSs]{protoneutron stars}
\newacro{RMNS}[RMNS]{remnant massive neutron star}
\acrodefplural{RMNS}[RMNSs]{remnant massive neutron stars}
\newacro{sGRBs}[sGRBs]{short gamma-ray bursts}
\newacro{SN}[SN]{supernova}
\newacro{SNR}[SNR]{signal-to-noise ratio}
\acrodefplural{SNR}[SNRs]{signal-to-noise ratios}
\newacro{TM}[TM]{tearing mode}
\newacro{TOV}[TOV]{Tolman-Oppenheimer-Volkoff}
\newacro{WENO}[WENO]{weighted essentially non-oscillatory}

\title[Subgrid modelling of MRI-driven turbulence in differentially rotating neutron stars]{Subgrid modelling of MRI-driven turbulence in differentially rotating neutron stars}

\author[M.~Miravet-Tenés et al.]{
Miquel Miravet-Tenés\orcidlink{0000-0002-8766-1156},$^{1,2}$\thanks{E-mail: m.miravet-tenes@soton.ac.uk}
Martin Obergaulinger\orcidlink{0000-0001-5664-1382},$^{2}$
Pablo Cerdá-Durán\orcidlink{0000-0003-4293-340X},$^{2,3}$
José A.~Font\orcidlink{0000-0001-6650-2634},$^{2,3}$
Milton Ruiz\orcidlink{0000-0002-7532-4144}$^{2}$
\\
$^{1}$Mathematical Sciences and STAG Research Centre, University of Southampton, Southampton SO17 1BJ, United Kingdom\\
$^{2}$Departament d'Astronomia i Astrofísica, Universitat de València, C/ Dr Moliner 50, 46100, Burjassot (València), Spain\\
$^{3}$Observatori Astronòmic, Universitat de València, C/ Catedrático José Beltrán 2, 46980, Paterna (València), Spain
}

\date{Accepted XXX. Received YYY; in original form ZZZ}

\pubyear{\the\year{}}

\begin{document}
\label{firstpage}
\pagerange{\pageref{firstpage}--\pageref{lastpage}}
\maketitle

\begin{abstract}
Following a binary neutron 
star (BNS) merger, the transient remnant is often a fast-spinning, differentially rotating,  magnetised hypermassive neutron star (HMNS).
This object is prone to the magnetorotational instability 
(MRI) which drives magnetohydrodynamic turbulence that significantly influences the HMNS global dynamics. A key consequence of turbulence is the outward transport of angular momentum which impacts the remnant’s stability and lifetime. Most numerical simulations of BNS mergers are unable to resolve the MRI due to its inherently small wavelength. To overcome this limitation, subgrid models have been proposed to capture the effects of unresolved small-scale physics in terms of large-scale quantities. We present the first implementation of our MHD-Instability-Induced Turbulence (MInIT) model in global Newtonian simulations of MRI-sensitive, differentially rotating, magnetised neutron stars. Here, we show that by adding the corresponding turbulent stress tensors to the momentum equation, MInIT successfully reproduces the angular momentum transport in neutron stars driven by small-scale turbulence.
\end{abstract}




\section{Introduction}

Multimessenger observations of \ac{BNS} mergers provide the most direct evidence that stellar compact mergers, where at least one of the binary companions is a \ac{NS}, may be progenitors of the central engines that power \acp{GRB}~\citep{MacFadyen:1999,GBM:2017lvd,gwgrb170817,Ruiz:2016}. They also give strong observational support to theoretical proposals linking \ac{BNS} mergers with production sites for $r$-process nucleosynthesis and kilonovae~\citep{1989Nature_nucleo,Li:1998bw,Metzger:2010}. Moreover, they can be used as standard sirens to give an independent measure of the expansion of the Universe~\citep{Schutz:1986,Nissanke:2010,Abbott:2017hubble}, and help put tight constraints on the \ac{EOS} of matter at supranuclear densities~\citep[see, e.g.,][and references therein]{Margalit:2017dij,Shibata:2017xdx,Rezzolla:2017aly,Ruiz:2017due}.

After merger, the system settles down into a new configuration. The merger outcome strongly depends on the total mass of the system and on the \ac{EOS} considered~\citep[see, e.g.,][for reviews]{Piro:2017,Bernuzzi:2020,Sarin:2021}. If the total mass of the remnant is somewhat larger than the mass of a stationary non-rotating \ac{NS} (\ac{TOV} solutions with mass $M_{\rm TOV}$), the system may go through a phase in which a transient post-merger object forms, a so-called \ac{HMNS}, supported against gravitational collapse by rapid differential rotation and thermal pressure\footnote{Temperatures in \ac{BNS} mergers may be  $\sim 100$ MeV and the inclusion of thermal effects in the \ac{EOS} is 
needed~\citep{Perego:2019,Hammond:2021}.}. The maximum mass of these remnants depends on the \ac{EOS}~\citep{Baumgarte:2000,Shibata:2006,Bauswein:2013b,Piro:2017,Weih:2018,Espino:2019}. The \ac{HMNS} may survive for several tens (or even hundreds) of milliseconds, undergoing oscillations and \ac{MHD} instabilities, and ejecting mass that forms a disc around the bulk of the star. Both the rotational profile and the disc mass depend on the \ac{EOS}~\citep[e.g.,][]{Kastaun:2015} and the mass ratio of the binary system~\citep[e.g.,][]{Bernuzzi:2020}. Massive NS remnants, apart from being differentially rotating, are characterised by strong magnetic fields~\citep[up to $B \sim 10^{16}$G, e.g.,~][]{Kiuchi:2014,Palenzuela:2022}. Such large values are the result of turbulent amplification periods both during and after merger, due to \ac{MHD} instabilities such as the \ac{KHI}, when the \acp{NS} are merging, and the \ac{MRI}, during the post-merger phase~\citep[e.g.,][]{Duez:2006b,Anderson:2008,Liu:2008,Kiuchi:2014,Kiuchi:2015,Kiuchi:2018,Kiuchi:2024,Ruiz:2016,Kawamura:2016, Palenzuela:2022}. 

Once support against gravity by rapid rotation and thermal pressure diminishes, the remnant eventually collapses to a black hole.  Damping of differential rotation comes from magnetic and viscous dissipation, i.e.,~angular momentum transport, that may arise from instabilities such as the \ac{MRI}~\citep{Balbus:1998,Duez:2004,Duez:2020,Siegel:2013,Radice:2018,Margalit:2022}. It is also worth mentioning that the stability of protoneutron stars, which also exhibit rapid differential rotation at birth, can be also influenced by the development of the \ac{MRI}~\citep[e.g.,][]{Akiyama:2003,Obergaulinger:2006b,Cerda:2008,Rembiasz:2016a,Reboul-Salze:2021}. Ionised rotating fluids with angular frequency profiles decreasing outwards are particularly unstable to the \ac{MRI}~\citep{Velikhov:1959,Chandrasekhar:1960,Balbus:1991} when threaded by a weak magnetic field in the direction perpendicular to the shear. Seed perturbations can grow exponentially on timescales close to the rotational period. These perturbations take the form of so-called ``channel modes'', which are pairs of vertically stacked layers in which the velocity and the magnetic field perturbations have radial and azimuthal components of (sinusoidally) alternating polarity. These modes have associated Maxwell and Reynolds stresses that lead to outward transport of angular momentum~\citep{Goodman:1994,Pessah:2006a,Pessah:2008}. The \ac{MRI} possesses a critical wavelength, $\lambda_{\rm MRI} \approx 2\pi v_{\rm A}/\Omega$~\citep[e.g.,][]{Shibatabook:2015}, which scales with the Alfvén speed $v_{\rm A}$ and the rotation frequency of the fluid $\Omega$, and corresponds to the fastest-growing mode. In the context of \ac{BNS} mergers, simulations focus on solving this mode~\citep{Siegel:2013, Kiuchi:2018, Kiuchi:2024,Ciolfi:2019}. However, this spatial scale is typically of the order of only tens of meters, making it challenging to resolve the MRI in numerical simulations of BNS systems. 

The exponential growth of the instability eventually terminates. The laminar \ac{MRI} channel flows can be unstable against \acp{PI}~\citep{Goodman:1994,Lesaffre:2009,Latter:2009,Miravet:2024b} that can be of KH or tearing mode type, depending on the value of kinematic viscosity and resistivity, i.e.,~non-ideal effects~\citep{Pessah:2009,Pessah:2010}. These secondary instabilities initially grow slowly, but eventually they evolve faster than the \ac{MRI} modes, since their growth rate is exponential to the \ac{MRI} amplitude. When both primary and secondary instabilities reach a similar amplitude, the channel modes are disrupted and the \ac{MRI} saturates~\citep{Rembiasz:2016a,Rembiasz:2016b}, leading to a turbulent regime. 

Numerical simulations of astrophysical systems such as BNS mergers, \ac{NSBH} mergers, and core-collapse supernovae are inherently challenging due to the complex and multifaceted physics involved. One key issue is capturing small-scale turbulence \citep[e.g.,][]{Radice:2024}. The prohibitive spatial resolution required to resolve all scales  prevents \ac{GRMHD} simulations from properly describing the turbulence triggered by \ac{MHD} instabilities. An emerging alternative is the use of \ac{LES}, which have already been employed to simulate both \ac{BNS} and \ac{NSBH} mergers~\citep{Giacomazzo:2015,Radice:2020,Aguilera-Miret:2022,Palenzuela:2022,Izquierdo:2024}. This approach aims to model, through the application of a subgrid closure, the small-scale turbulence in terms of resolved quantities. More precisely, LES provide a closure for the turbulent stress tensors, which appear in the nonlinear mean-field \ac{MHD} equations. 

In~\citet{Miravet:2022,Miravet:2024}, we presented a new Newtonian subgrid model for \ac{MHD} turbulence triggered by the \ac{MRI} and the \ac{KHI}, the dominant \ac{MHD} instabilities in \ac{BNS} mergers. The model, dubbed \ac{MInIT}, is based on evolution equations for the turbulent kinetic energy densities. These equations are built using phenomenological arguments that are physically motivated. The turbulent densities are connected to the stress tensors through certain calibrated coefficients. This model allows handling delays in the growth of the instability and the decay of turbulence, and it has been calibrated by fully resolved local numerical simulations. Moreover, it has been adapted to the instabilities that are key drivers of turbulence in BNS mergers, in contrast to other models already applied to \ac{LES}, which are only able to partially capture the magnetic field amplication driven by the \ac{KHI}, but do not show any evidence of \ac{MRI} development. The gradient model employed in, e.g.,~\citet{Palenzuela:2022} and~\citet{Aguilera-Miret:2025}, seems to provide promising results when dealing with partially resolved turbulence, as in the case of the \ac{KHI}, since the model seems to extrapolate the turbulent cascade to the unresolved small scales. However, there is no evidence that this model is able to capture the impact of subgrid turbulence triggered by the \ac{MRI}, because this instability is expected to fully develop in the unresolved scales.

In this work, we use the \ac{MInIT} model in global Newtonian simulations of MRI-sensitive, differentially rotating, magnetised \acp{NS} and evaluate its capability to accurately resolve the MRI dynamics. We focus on the angular momentum transport arising from the inclusion of subgrid terms in the momentum equation, deferring to a future work the effect of the subgrid scales on the expected magnetic field amplification after \ac{MRI} saturation. By exploring different rotational frequencies, magnetic field strengths, and initial values of the turbulent energy densities, we study their impact on the angular momentum transport timescale in simulations that lack enough spatial resolution to directly resolve the \ac{MRI}. 

This paper is organised as follows: in Section~\ref{sec::mean_field_eqs} we introduce the mean-field \ac{MHD} equations with the inclusion of the turbulent stresses. We describe in Section~\ref{sec::minit} the closure model for turbulence we employ in the simulations. The numerical methodology is discussed in Section~\ref{sec::methods} and the results are showcased in Section~\ref{sec::results_sims}. Conclusions are drawn in Section~\ref{sec::conclusions}, along with prospects for future research. Finally, Appendix~\ref{app1} discusses the effect numerical dissipation might have in our simulations. Unless otherwise stated we employ cgs units. Latin indices run from 1 to 3. 

\section{Mean-field MHD equations}
\label{sec::mean_field_eqs}

We start by briefly reviewing the Newtonian ideal MHD equations which form the mathematical framework for our study. These equations couple the different variables of a plasma, such as the gas pressure, the mass density, the velocity and the magnetic field. We can express this system of equations as
\begin{eqnarray}
    \partial_t \rho + \nabla_j \big[\rho v^j \big] &=& 0, \label{rho_eq}\\
    \partial_t p^i+\nabla_j \big[ p^i v^j+P_{\star}\delta^{ij} -b^ib^j \big] &=& f^i, \label{press_eq} \\
    \partial_t e_{\star} +\nabla_j \big[ (e_{\star}+P_{\star} ) v^j-b^iv_ib^j\big] &=& f^jv_j, \label{ener_eq} \\
    \partial_t \vec{b} &=& -c \vec{\nabla} \times \vec{E}, \label{ind_eq} \\
    \nabla_j b^j &=& 0 \label{div_eq} \,,
\end{eqnarray}
where $\rho$ is the mass density, $v^i$ are the components of the fluid velocity, $b^i$ are the magnetic field components, $p^i = \rho v^i$ is the momentum density, $P_{\star} = P_{\rm gas}+b^2/2$ is the total pressure, $e_{\star} = e_{\rm int}+\rho v^2/2+b^2/2$ is the total energy density, and $f^i$ is an external force density, which, in this case, corresponds to gravity, $f_i = -\rho \nabla_i\Phi$. The gravitational potential $\Phi$ is computed as
\begin{equation}\label{newt_potential}
    \Phi (r) = -4\pi \int_0^{\infty}dr' r'^2\frac{\rho}{|r -r'|}\,,
\end{equation}
where $r$ is the radial spherical coordinate. By applying the mean-field \ac{MHD} formalism~\citep{krause,Miravet:2022,Miravet:2024}, the above system of equations can be expressed in terms of resolved and subgrid-scale terms. If we assume that the behaviour of a given field $\boldsymbol{A}$ is solved for a certain lengthscale $l$, we can introduce a filter that acts on that scale. The residual between filtered and unfiltered fields will be the turbulent contribution,
\begin{equation}
    \boldsymbol{A'} = \boldsymbol{A}-\bar{\boldsymbol{A}}\,,
\end{equation}
where the bar symbol 
denotes the filtering/averaging operator and the prime symbol is used to  
identify the turbulent field. This filtering operation satisfies the Reynolds averaging rules and can be either temporal or spatial~\citep{book}. By introducing this decomposition in the \ac{MHD} equations, which are nonlinear, and after applying the filtering operation to the equations, additional terms with products of turbulent quantities will appear. By construction, the average of the turbulent contribution is zero. Thus, the only possible turbulent terms arising in the mean-field equations are the averaged products of two or more unresolved variables. Following~\citet{Miravet:2022,Miravet:2024} we  consider only  fluctuations of the velocity and magnetic fields. Therefore, the average of the products of these unresolved variables can be represented by
\begin{eqnarray}\label{stresses}
    \bar{M}_{ij} &=& \overline{b'_i b'_j},\\
    \bar{R}_{ij} &=& \overline{v'_i v'_j}, \\
    \bar{F}_{ij} &=& \overline{v'_i b'_j}-\overline{v'_j b'_i}\,,
\end{eqnarray}
which correspond to the Maxwell, Reynolds and Faraday turbulent stress tensors, respectively. Linear combinations of these terms will appear as effective source terms in the mean-field version of the system~\eqref{rho_eq}-\eqref{div_eq}.

Since the aim of this work is to solely study angular momentum transport, we will focus on the mean-field form of the momentum equation,
\begin{equation}\label{momentum_filt}
    \partial_t \bar{p}^i+\nabla_j \big[ \bar{\rho} \bar{v}^i \bar{v}^j+\big(\bar{P}_{\star}+\Tr{\bar{\boldsymbol{M}}}\big)\delta^{ij} -\bar{b}^i\bar{b}^j +\bar{\rho}\bar{{R}}^{ij}-\bar{M}^{ij} \big] = \bar{f}^i\,,
\end{equation}
where the trace of the Maxwell stress can be regarded as a turbulent magnetic pressure. Density perturbations are neglected, which leaves the continuity equation unchanged. For the sake of simplicity we do not include any subgrid term in the energy and induction equations. Therefore, we do not expect a turbulent dynamo that exponentially amplifies the large-scale magnetic field. A study of the effect of the \ac{MRI} turbulent dynamo in the large-scale dynamics is deferred to future work.

\section{The MInIT model for the MRI}
\label{sec::minit}

The turbulent stress tensors that appear in Eq.~\eqref{momentum_filt} need a closure relation, i.e., they need to be connected to the resolved variables in order to write the system of equations as a closed system amenable to be solved numerically. 
In the \ac{MInIT} subgrid model~\citep{Miravet:2022} the closure relation is obtained by introducing a new quantity, the turbulent kinetic energy density, with an evolution equation of the form
\begin{equation}
    \partial_t e_{\rm turb} + \nabla_j(\bar{v}_j e_{\rm turb}) = S^{\rm turb}\,, 
\end{equation}
where $S^{\rm turb}$ comprises source terms that depend on the specific kind of \ac{MHD} turbulence under consideration. In the context of this work, the dominant \ac{MHD} instability, and the one that  will develop subgrid turbulence, is the \ac{MRI}. As shown in~\citet{Miravet:2022}, the secondary \acp{PI} are responsible for the saturation of the \ac{MRI}. Therefore, we need two evolution equations to account for the two instabilities, the \ac{MRI} and the \ac{PI}: 
\begin{eqnarray}
    \partial_t e_{\rm MRI} + \nabla_j (\bar{v}_j e_{\rm MRI})&=& 2\,\gamma_{\rm MRI}\,e_{\rm MRI}-2\,\gamma_{\rm PI}\,e_{\rm PI}\,,\label{turb_en_mri} \\
    \partial_t e_{\rm PI} + \nabla_j (\bar{v}_j e_{\rm PI})&=&  2\,\gamma_{\rm PI}\,e_{\rm PI}-S_{\rm TD}\,. \label{turb_en_pi}
\end{eqnarray}

In the ideal \ac{MHD} case, the explicit form of the \ac{MRI} growth rate of the fastest-growing mode, $\gamma_{\rm MRI}$, is~\citep{Balbus:1995,Obergaulinger:2009}
\begin{equation}\label{grate_mri}
    \gamma_{\rm MRI} = \frac{q}{2}\Omega\,,
\end{equation}
where $\Omega$ is the angular frequency of the fluid and $q$ is known as the shear parameter
\begin{equation}\label{shear}
    q \equiv - \frac{d \ln \Omega}{d \ln r}\,.
\end{equation}

Correspondingly, the growth rate of the \acp{PI} can be expressed as~\citep{Pessah:2010,Miravet:2022}
\begin{equation}\label{grate_pi}
    \gamma_{\rm PI} = \sigma k_{\rm MRI}\sqrt{\frac{2 e_{\rm MRI}}{\rho}}\,,
\end{equation}
with $\sigma = 0.27$~\citep{Pessah:2010} and $k_{\rm MRI}$ being the wavenumber of the fastest-growing \ac{MRI} mode~\citep{Rembiasz:2016a},
\begin{equation}\label{wavenumber}
    k_{\rm MRI} = \sqrt{1-\frac{(2-q)^2}{4}}\frac{\Omega}{\bar{v}_{\rm Az}}\,,
\end{equation}
where $\bar{v}_{\rm Az} = \bar{b}_z/\sqrt{\rho}$ is the vertical component of the Alfvén velocity. In practice, since the vertical and poloidal components of the magnetic field are very similar in our simulations, we will use the latter to avoid divisions by zero at certain points of the domain. The growing term for the \ac{PI} in Eq.~\eqref{turb_en_pi}, i.e., the source term with positive sign, acts as a sink for the \ac{MRI} energy in Eq.~\eqref{turb_en_mri}, since the secondary instabilities feed off the main one.    The sink term from Eq.~\eqref{turb_en_pi}, $S_{\rm TD}$, represents the dissipation of the turbulent kinetic energy into thermal energy at the end of the Kolmogorov scale, i.e., the inertial range of scales~\citep{Landau:1987,Miravet:2022}. It is given by
\begin{equation}\label{kolmogorov}
    S_{\rm TD} = C \frac{e_{\rm PI}^{3/2}}{\sqrt{\rho}\lambda}\,,
\end{equation}
where $C = 8.6$, a value empirically found in~\citet{Miravet:2022}, and $\lambda = \min[\Delta,\lambda_{\rm MRI}]$, with $\Delta$ the numerical cell size and $\lambda_{\rm MRI}$ the wavelength of the fastest growing \ac{MRI} mode, $\lambda_{\rm MRI} = 2\pi/k_{\rm MRI}$.

The stress tensors are linked to these turbulent energy densities through constant proportionality coefficients,
\begin{eqnarray}
    \bar{M}_{ij}(t,\boldsymbol{r}) & = &\alpha^{\rm MRI}_{ij} \,e_{\rm MRI}(t,\boldsymbol{r}) + \alpha^{\rm PI}_{ij} \,e_{\rm PI}(t,\boldsymbol{r})\,, \label{Maxwell_coeff}\\
	\bar{R}_{ij}(t,\boldsymbol{r}) & = &\frac{1}{\bar{\rho}(t, \boldsymbol{r})} \left ( \beta^{\rm MRI}_{ij} \,e_{\rm MRI}(t,\boldsymbol{r}) + \beta^{\rm PI}_{ij} \,e_{\rm PI}(t,\boldsymbol{r}) \right)\,, \\
	\bar{F}_{ij}(t,\boldsymbol{r}) & = &\frac{\gamma^{\rm PI}_{ij}}{\sqrt{\bar{\rho}(t,\boldsymbol{r})}} e_{\rm PI}(t,\boldsymbol{r}) \,, 
\end{eqnarray}
which are either obtained from theoretical arguments~\citep[in the case of the \ac{MRI} coefficients;][]{Pessah:2008} or calibrated using numerical box simulations~\citep[for the \ac{PI} coefficients;][]{Miravet:2022}. The dominant contribution responsible for angular momentum transport in the momentum equation are the cylindrical $\varpi\phi$ components of the Maxwell and Reynolds stresses. Here, the quantity $\varpi$ corresponds to the cylindrical radius, i.e., the distance to the rotation axis, $\varpi = r\sin\theta$. The coefficients corresponding to the MRI components are
\begin{eqnarray}
    \alpha^{\rm MRI}_{\varpi\phi} &=& 1-4/q\,, \label{alpha_MRI}\\
    \beta^{\rm MRI}_{\varpi\phi}&=& 1\,,  \label{beta_MRI}
\end{eqnarray}
while the calibrated parasitic coefficients are $\alpha^{\rm PI}_{\varpi\phi}= -1.4$ and $\beta^{\rm PI}_{\varpi\phi} = -0.8$. Uncertainties (standard deviation) in these quantities arise from both the spatial and time averages performed over the simulation data in~\citet{Miravet:2022}. The rest of the coefficients can be found in~\citet{Miravet:2022}. 

\section{Numerical approach}
\label{sec::methods}

\subsection{Initial models}
\label{subsec::id}

The differentially rotating equilibrium models are computed using the Newtonian version of the code described by~\citet{Dimmelmeier:2002bk}, based on Hachisu's self-consistent field method~\citep{Komatsu:1989a}. The rotation law of the equilibrium model is given by
\begin{equation}\label{rot_law}
    \Omega (\varpi) = \frac{\Omega_c}{1+\frac{\varpi^2}{A^2}}\,,
\end{equation}
where $A$ is a positive constant and $\Omega_c$ is the value of $\Omega$, the angular frequency, at the coordinate centre~\citep{Komatsu:1989a}. In the limit where $A\rightarrow \infty$, the star becomes a rigid rotator. The initial values of the turbulent energy densities of the \ac{MRI} and the \ac{PI} will be a fraction of the total kinetic energy density (see Sect.~\ref{subsec::turb_amp}). 

Regarding the \ac{EOS}, a polytropic relation between the pressure $P$ and the rest-mass density $\rho$ is employed:
\begin{equation}\label{poly_eos}
    P = K \rho^{\gamma}\,,
\end{equation}
with $\gamma = 2$ and $K = 145529.19$ g$^{-1}$ cm$^{5}$ s$^{-2}$. 

A dipolar magnetic field is implemented as in~\citet{Suwa:2007}, with the following components of the effective vector potential (in a spherical coordinate system),
\begin{eqnarray}\label{suwa_dipole}
    A_r & = & 0 \,, \\
    A_{\theta} & = & 0 \,, \\ 
    A_{\phi} & = & \frac{\bar{b}_0}{2}\frac{r_0^3}{r^3+r_0^3}\varpi \times\rm{max}(0,(\rho-\rho_{\rm cut})/ \rho_{\rm max})\,, 
\end{eqnarray}
where $r_0$ and $\bar{b}_0$ are model constants, the latter being the value of the magnetic field at the center of the star. In all our initial models we 
set $r_0 = 12$ km, being the equatorial radius of the star $R_{\rm eq} \approx 18.5$ km (see Table~\ref{table:omegas}). The last factor in the expression for $A_{\phi}$ is included to keep (initially) the magnetic field confined inside the star~\citep{Etienne:2012,Ruiz:2021}. The cutoff density $\rho_{\rm cut}$ is a free parameter that confines the magnetic field within $\rho > \rho_{\rm cut}$. We set $\rho_{\rm cut}$ to $10\%$ of the initial maximum density, which corresponds to an equatorial radial distance of roughly 17 km for all simulations. 

\begin{figure}
    \centering
    \includegraphics[width=\linewidth]{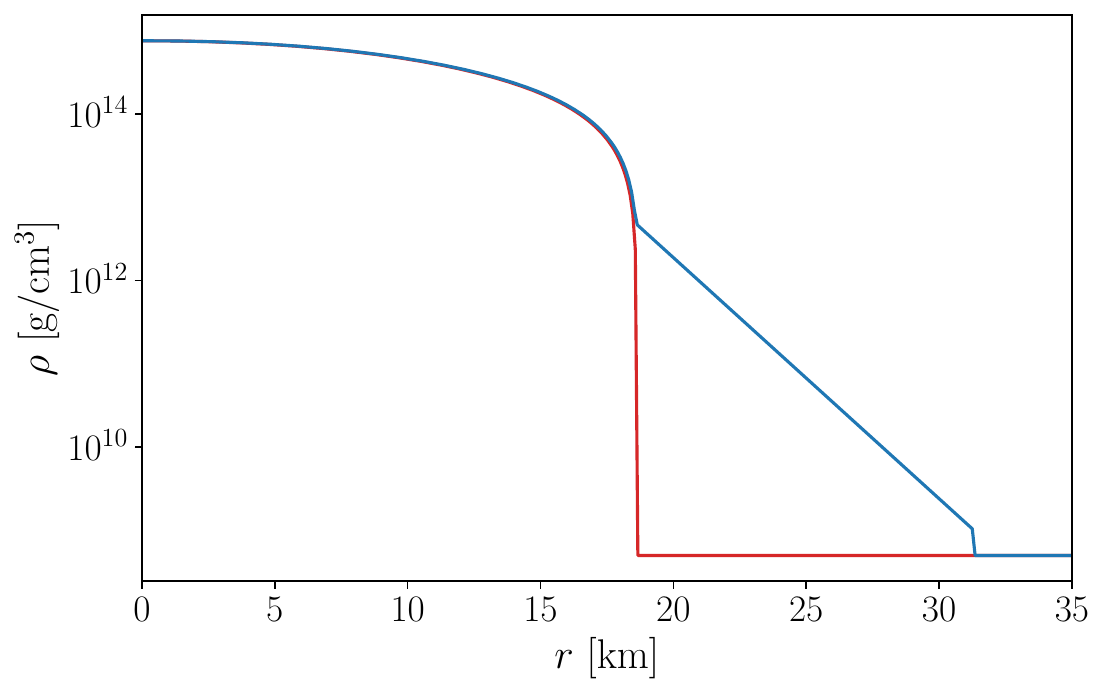}
    \caption{Radial equatorial profile of the initial mass density with (blue) and without (red) the exponential decay at low values. The profiles are identical for density values above $\rho_{\rm thresh, atm}$.}
    \label{fig:rho_prof}
\end{figure}

The polytropic EOS in Eq.~\eqref{poly_eos} leads to a very steep density profile at the surface of the neutron star. In order to deal with the vacuum region surrounding the star we use a low-density atmosphere, as customary in these kind of simulations, where the hydrodynamical variables are not evolved. The threshold value of the mass density to characterise the atmosphere is set to $\rho_{\rm atm} = 10^9$ g/cm$^3$, which is roughly five orders of magnitude smaller than the initial central density of the star (see Table~\ref{table:omegas}). The rapid decrease of the density with the radial distance can lead to numerical instabilities at the interphase between the star and the atmosphere. In order to prevent that from happening, we add an exponential radial decay of the density profile that smooths the transition to the atmosphere, for densities below a given threshold value $\rho_{\rm thresh, atm}$,
\begin{equation}
    \rho(r.\theta) = {\rm max}\Big(\rho_{\rm thresh,atm}\exp\Big[ \frac{r-r_{\rm atm}(\theta)}{\Delta_r}\Big], \rho_{\rm atm}\Big)\,.
\end{equation}
Both $\rho_{\rm thresh,atm}$ and $\Delta_r$ are freely specifiable parameters. The quantity $r_{\rm atm}$ is the radius at which $\rho = \rho_{\rm thresh,atm}$. Fig.~\ref{fig:rho_prof} shows this density profile as compared to that without the exponential decay (indicated by the red curve). Table~\ref{table:omegas} summarises our sample of initial models, reporting the maximum mass density, $\rho_{\rm max}$, the equatorial radius, $R_{\rm eq}$, the central rotation frequency, $\Omega_c$, the total angular momentum, $J$, and the strength of the initial poloidal magnetic field, $\bar{b}_0$. For all models we fix the gravitational mass of the star to $M_{\rm grav}=2.60$ $M_{\odot}$. Moreover, the value of $\rho_{\rm thresh,atm}$ that we employ corresponds to $5\times 10^{12}$ g/cm$^{-3}$, which is more than two orders of magnitude smaller than $\rho_c$.

\begin{table*}
 \caption{Summary of the initial models. All stars have the same gravitational mass $M_{\rm grav}=2.60$ $M_{\odot}$. The columns report the label of the model, the maximum mass density, $\rho_{\rm max}$, the equatorial radius, $R_{\rm eq}$, the central rotation frequency, $\Omega_c$, the total angular momentum, $J$, and the strength of the initial poloidal magnetic field, $\bar{b}_0$.}
    \centering 
    \begin{tabular}{cccccc}
        \toprule[0.5pt]
        \rule{0pt}{10pt} LABEL  &  $\rho_{\rm max}$ & $R_{\rm eq}$ & $\Omega_c$  & $J$  & $\bar{b}_0$  \\ 
        \rule{0pt}{10pt}   &  [$10^{14}$ g/cm$^3$] & [km] &[s$^{-1}$] & [$10^{48}$ g cm$^2$/s] & [$10^{13}$ G]
        \\ \midrule[0.5pt]
         \rule{0pt}{10pt}{$\Omega_1$} &  7.90 & 18.53 & 2887.85 & 3.12 & 7.00  \\
         \rule{0pt}{10pt}{$\Omega_2$} &  7.80 & 18.58 & 4922.76 & 5.32 & 7.00 \\
         \rule{0pt}{10pt}{$\Omega_3$} &  7.61 & 18.70 &  7500.03 & 8.12  & 7.00  \\
         \rule{0pt}{10pt}{$\Omega_3$-b1e14} &  7.61& 18.70 & 7500.03 & 8.12  & 10.00  \\
         \rule{0pt}{10pt}{$\Omega_3$-b8.5e13} &  7.61& 18.70 & 7500.03 & 8.12  & 8.50   \\
         \rule{0pt}{10pt}{$\Omega_3$-b5e13} &  7.61& 18.70 & 7500.03 & 8.12  & 5.00   \\
         \rule{0pt}{10pt}{$\Omega_3$-b3.5e13} &  7.61& 18.70 & 7500.03 & 8.12  & 3.50   \\
         \bottomrule[0.5pt]
    \end{tabular}
    \label{table:omegas}
\end{table*}

\subsection{Numerical evolution}
\label{subsec::code}

\subsubsection{General considerations}

The initial models reported in Table~\ref{table:omegas} are evolved using the \textsc{Aenus} code~\citep{Obergaulinger-2008} which solves the ideal \ac{MHD} equations in their conservative form using finite-volume methods. The simulations are performed using the \ac{HLL} flux formula~\citep{Harten:1983}, a \ac{PPM} reconstruction for cell interfaces~\citep{ColW84}, and a $3^{\mathrm{rd}}$ order Runge-Kutta time integrator~\citep{Shu:1988}. For the spatial grid, the code employs spherical polar coordinates $(r,\theta,\phi)$ and axial symmetry with respect to the rotation axis is assumed. The number of grid cells is $(N_r,N_{\theta}, N_{\phi}) = (468,180,1)$, with $r \in [0,50]$ km and $\theta \in [0,\pi]$ rad. Since the \ac{MInIT} coefficients are computed in cylindrical coordinates in~\cite{Miravet:2022}, a change of basis from cylindrical to spherical coordinates is required. Furthermore, the angular frequency $\Omega$, and the shear $q$, are computed from the angular velocity $v_{\phi}$. For the radial direction we use boundary conditions that ensure regularity at the geometric singularity of the origin and employ a constant extrapolation at the outer edge of the grid. Regarding the polar direction, conditions adapted to the polar axis are used. 

\subsubsection{Implementation of the MInIT model}

In order to solve Eqs.~\eqref{turb_en_mri} and~\eqref{turb_en_pi}, the code employs high-resolution shock-capturing schemes for the transport terms on the left-hand side, as done for the other \ac{MHD} quantities. As the source terms can be stiff, we treat them in an operator-split manner using an implicit integration, similarly to what~\citet{Just:2015b} did to deal with neutrino schemes.

After the transport of angular momentum, the shear parameter $q$ tends to zero in certain regions of the star. Moreover, there might be regions that are not unstable to the MRI, which correspond to the cases with $q < 0$ or $q > 4$~\citep{Obergaulinger:2009}. For $q<0$ the fluid is stable and for $q>4$ the fluid is subject to large scale shear instabilities, that can be resolved numerically without the need of a subgrid model. As can be seen from Eq.~\eqref{wavenumber}, the \ac{MRI} wavenumber becomes imaginary for those values of $q$. For $q \rightarrow 0$ and $q \rightarrow 4$, the \ac{MRI} wavelength tends to infinity. This implies that at some value of $q$ close enough to $0$ or $4$, $\lambda_{\rm MRI} > \Delta$. In such cases, $\lambda_{\rm MRI}$ is replaced by $\Delta$ and therefore we substitute $k_{\rm MRI}$ in Eq.~\eqref{grate_pi} by $k = 2\pi / \Delta$, as done for the turbulent Kolmogorov term in Eq.~\eqref{kolmogorov}. Moreover, since the \ac{MRI} is not expected to develop for $q \leq 0$ and $q \geq 4$, we set $\gamma_{\rm MRI}$ to zero in those cases.

The flattening of the rotational profile of the star as a result of angular momentum transport can also lead to some issues when computing the \ac{MRI} contribution of the Maxwell stress tensor, since the $\alpha^{\rm MRI}_{ij}$ coefficients depend on $1/q$. To solve this potential issue, we set $\alpha^{\rm MRI}_{ij}$ to zero when $q \leq 10^{-3}$. This should not be problematic, since we expect $e_{\rm MRI}$ and $e_{\rm PI}$ to decay when $q \rightarrow 0$. 

\section{Results}
\label{sec::results_sims}

Including turbulent energy densities in the simulations introduces an interplay with large-scale quantities, influencing the evolution of the two of them.
In the following we test the impact of different initial values of both large-scale and small-scale quantities on the evolution of the MRI and, consequently, on the large-scale dynamics of the star.

\subsection{Dependence on the central rotation frequency}
\label{subsec::omegas}

\begin{figure}
    \centering
    \includegraphics[width=\linewidth]{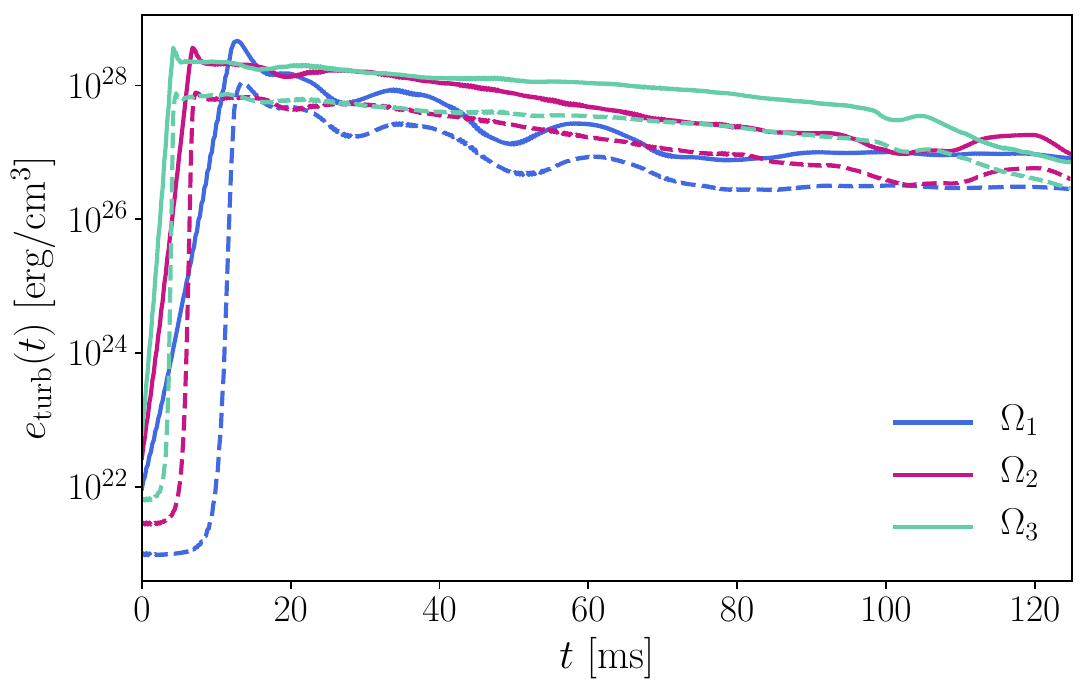}
    \caption{Time evolution of the turbulent energy densities, $e_{\rm MRI}$ (solid lines) and $e_{\rm PI}$ (dashed lines), averaged over a radius of $r = 4$ km. Different colours represent different central initial rotation frequencies, $\Omega_i$. }
    \label{fig:eturb_tev}
\end{figure}

From Eqs.~\eqref{grate_mri} and~\eqref{grate_pi} it follows that the rotation frequency $\Omega$ has a direct impact on the growth of the turbulent energy densities of the \ac{MInIT} model. Fig.~\ref{fig:eturb_tev} depicts the time evolution of the average  of these quantities over a sphere of radius $r = 4$ km, for different values of the initial central rotation frequency, given by the models $\Omega_1$, $\Omega_2$ and $\Omega_3$ from Table~\ref{table:omegas}. We choose such a low value of $r$ to also depict the turbulence decay, which mainly happens in the central regions of the star. The initial values of $e_{\rm MRI}$ and $e_{\rm PI}$ are chosen to be $10^{-10}e_{\rm kin}(0)$ and $10^{-11}e_{\rm kin}(0)$, respectively. We will explore in Sect.~\ref{subsec::turb_amp} the impact of different initial values of these quantities on the simulations. To evolve these quantities only at the stellar interior, we set $e_{\rm MRI}(0) = e_{\rm PI}(0) = 0$ for $\rho < 0.1\rho_{\rm max}$, which corresponds to a radius of $\sim 17$ km, as mentioned before.

As expected, models with larger central rotation frequency, i.e.,~larger angular momentum, show a more rapid growth of both the \ac{MRI} and \ac{PI} energy densities. However, the saturation amplitude of both quantities is approximately the same, regardless of the value of the rotation frequency. Thus, the only difference between the turbulent energy densities for different rotation velocities is the time at which they saturate, which may have an impact on the timescale of the angular momentum transport. Moreover, after $t \approx 40$ ms, all the saturated turbulent energy densities start decaying, due to the transport of angular momentum that stabilises the central region of the star against the \ac{MRI}. 

\begin{figure}
    \centering
    \includegraphics[width=\linewidth]{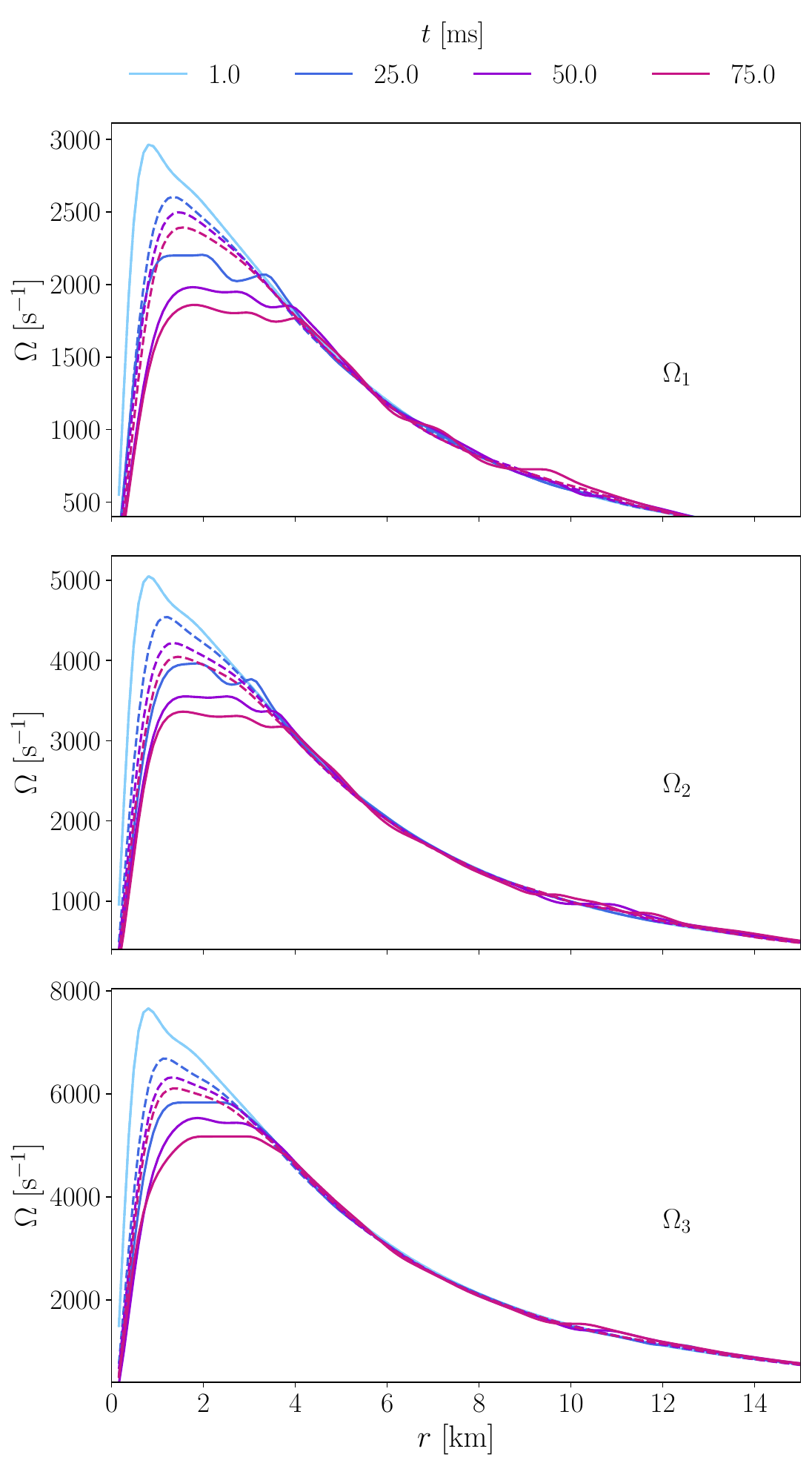}
    \caption{Radial equatorial profiles of the angular frequency, $\Omega$, for several initial central values, $\Omega_i$. Each colour represents different times, indicated in the top legend. 
    The solid curves correspond to simulations incorporating the \ac{MInIT} model while the dashed ones for simulations without it. For $t = 1$ ms, both solid and dashed curves overlap, because the \ac{MRI} is still developing.}
    \label{fig:omg_prof}
\end{figure}

The development of the \ac{MRI} has an impact on the large-scale dynamics of the neutron star. In Fig.~\ref{fig:omg_prof} we showcase the equatorial radial profiles of the angular frequency of the star. Solid lines correspond to simulations that incorporate the MInIT model while dashed lines to those not including it. Quite distinctly, \ac{MRI} turbulence leads to the flattening of the profile in the inner regions of the star, expanding outwards with time. Hence, the effect of the turbulent stress tensors in the momentum equation is the transport of angular momentum radially outwards. This is observed for all initial models of our sample with different rotation frequencies. Fig.~\ref{fig:omg_prof} also displays a slight increase of the angular frequency at larger radial distances, where angular momentum is being transported. We note that despite the dashed curves also show a decrease of the central angular frequency, this is purely due to numerical dissipation (see Appendix~\ref{app1} for details), and not to the effect of MRI-resolved turbulence. We have checked this by running a simulation with the same spatial resolution but without magnetic fields, observing the same behaviour on the rotational profile.
  
\begin{figure*}
    \centering
    \includegraphics[width=\linewidth]{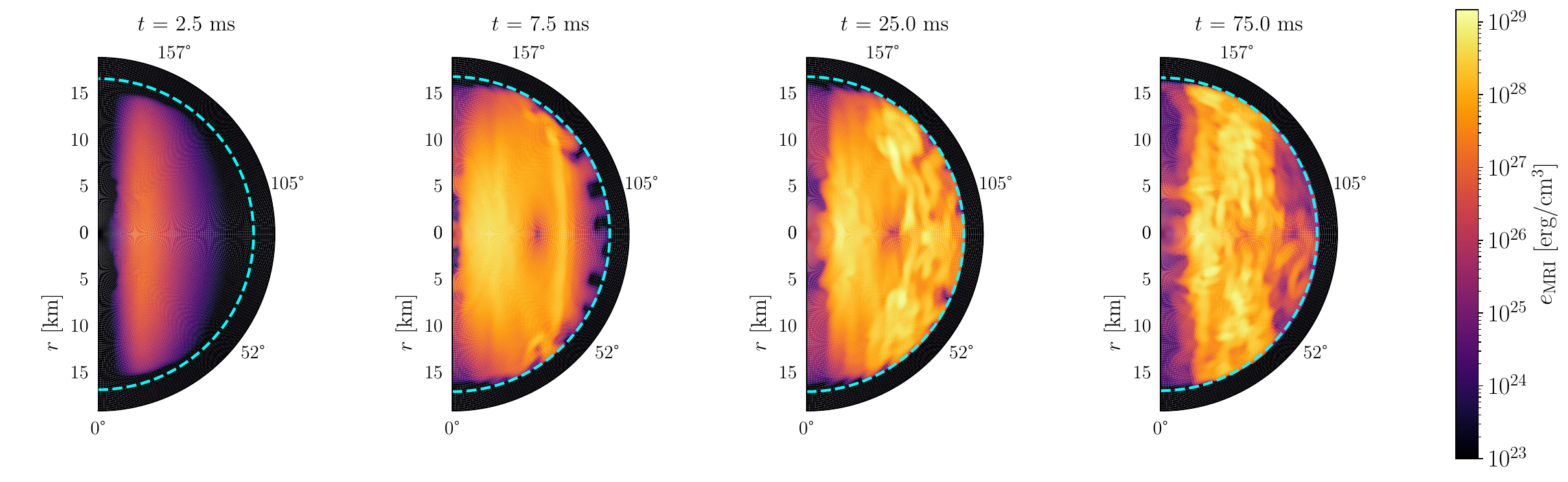}
    \caption{Contour plot of the turbulent kinetic energy density of the \ac{MRI}, $e_{\rm MRI}$. Each panel stands for different times $t = \{2.5,7.5,25,75\}$ ms. After a rapid growth during the first $\sim 10$ ms through all the stellar domain, the energy density starts decaying at the inner region of the star, due to the transport of angular momentum from the centre. The blue dashed line stands for the isocontour of the mass density $\rho$ at 10$\%$ of its central value.}
    \label{fig:eturb_prof}
\end{figure*}

\begin{figure}
    \centering
    \includegraphics[width=\linewidth]{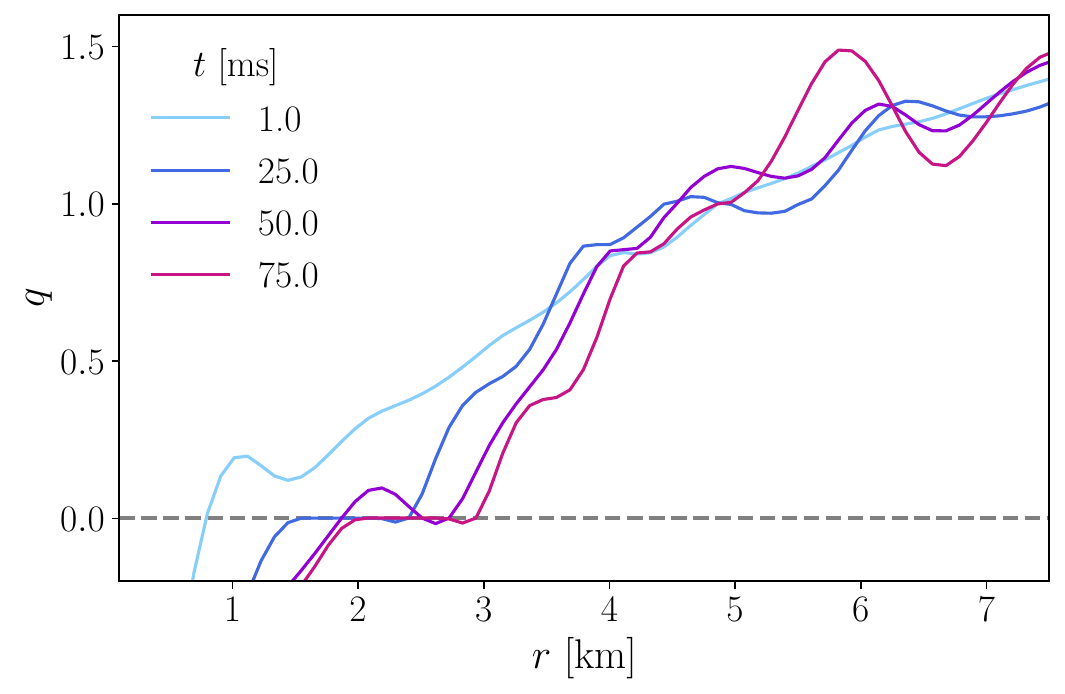}
    \caption{Radial equatorial profiles of the shear parameter at different times, corresponding to the simulation labelled with $\Omega_3$. Due to angular momentum transport, $q$ tends to 0 (dashed grey line) as time increases.}
    \label{fig:q_prof}
\end{figure}

Figure~\ref{fig:eturb_prof} shows the contour of the \ac{MRI} turbulent kinetic energy density, $e_{\rm MRI}$, for the model $\Omega_3$, at $t = \{2.5,7.5,25,75\}$ ms. There exists a rapid growth during the first 7.5 ms of the simulation, where $e_{\rm MRI}$ grows several orders of magnitude (from $\sim 10^{26}$ erg/cm$^3$ to $\sim 10^{29}$ erg/cm$^3$ for $\varpi$ up to $\sim 10$ km). After saturation, and when the redistribution of angular momentum sets in (see Fig.~\ref{fig:omg_prof}) $e_{\rm MRI}$ starts decaying at low values of the cylindrical radial coordinate, $\varpi$. The cylindrical symmetry is due to the choice of rotation law (see Eq.~\eqref{rot_law}), which depends on $\varpi$. The turbulence decay can be understood by looking at Fig.~\ref{fig:q_prof}, where radial profiles of $q$ are depicted for $t = \{1,25,50,75\}$ ms, as in Fig.~\ref{fig:omg_prof}. As the angular frequency profile flattens, the shear parameter is reduced until the region transitions to rigid rotation, i.e., $q \rightarrow 0$. When this happens, the \ac{MRI} is no longer active in that region and the turbulence generated by this instability can only decay.  

\subsection{Dependence on the magnetic field strength}
\label{subsec::bpol}

We also study the effect of the magnetic field strength on the evolution of the \ac{MRI} and the \ac{PI}, and on the global dynamics. We employ three different values of the central poloidal magnetic field, $\bar{b}_0 = \{3.5,7,10\}\times 10^{13}$ G, corresponding to the models $\Omega_3$-b3.5e13, $\Omega_3$ and $\Omega_3$-b1e14 from Table~\ref{table:omegas}, respectively. It is useful to study the effect of different initial poloidal fields on the evolution of the turbulent energy densities, since it explicitly appears in Eq.~\eqref{grate_pi}. 

\begin{figure}
    \centering
    \includegraphics[width=\linewidth]{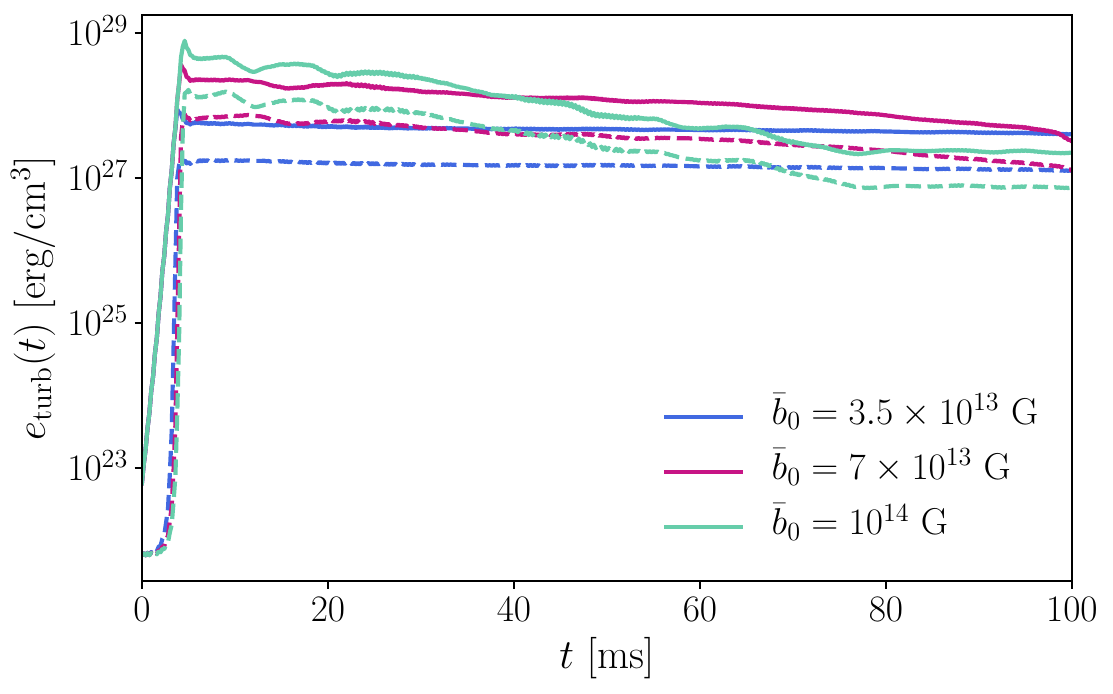}
    \caption{Time evolution of the turbulent energy densities, $e_{\rm MRI}$ (solid) and $e_{\rm PI}$ (dashed) averaged over a radius of $r = 4$ km. We show different choices for the initial poloidal magnetic field amplitudes at the centre of the star: $\bar{b}_0 = 3.5\times 10^{13}$ G (blue), $\bar{b}_0 = 7\times 10^{13}$ G (red), and $\bar{b}_0 = 10^{14}$ G (green). The central initial rotation frequency is fixed to $\Omega_3$.}
    \label{fig:eturb_tev_bcomp}
\end{figure}

\begin{figure}
    \centering
    \includegraphics[width=\linewidth]{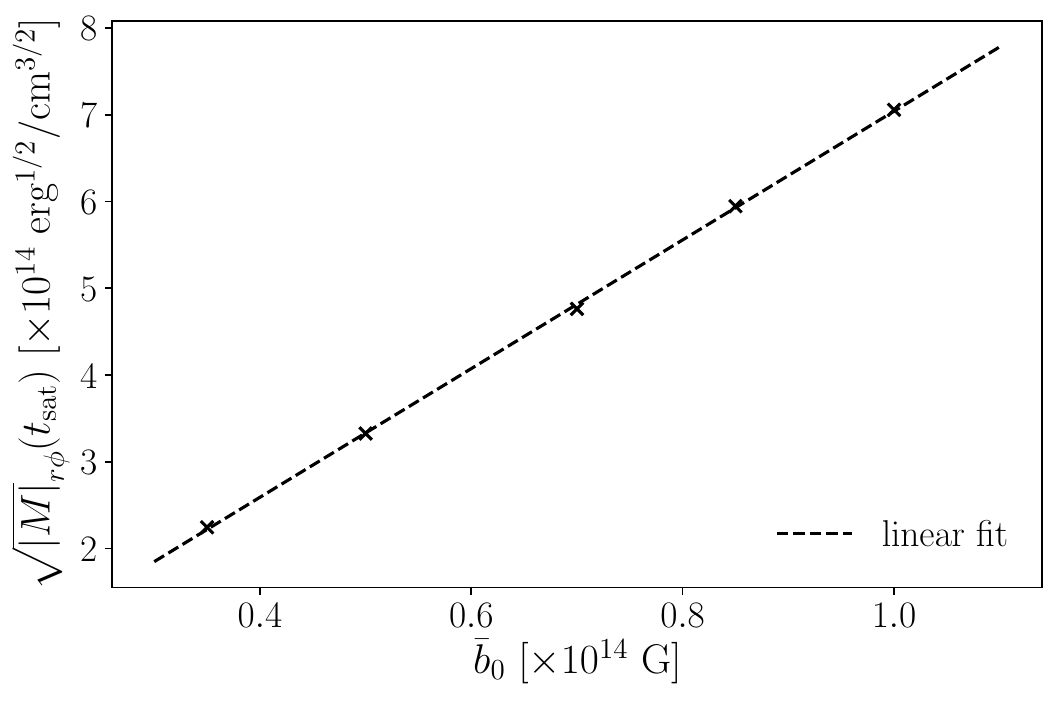}
    \caption{Absolute values at saturation of the average value of the square-rooted $r\phi$ component of the Maxwell stress tensor, over a radius of $r = 8$ km, as a function of the initial poloidal field amplitude. The dashed line represents the linear fit.}
    \label{fig:esat_vs_bpol}
\end{figure}

\begin{figure}
    \centering
    \includegraphics[width=\linewidth]{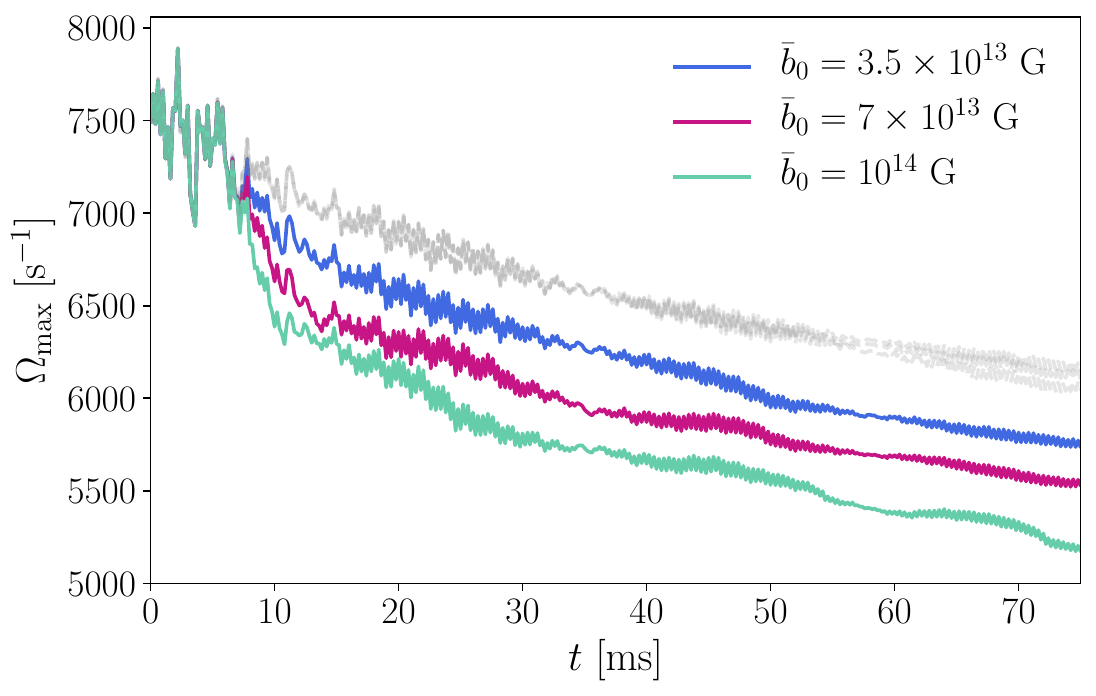}
    \caption{Time evolution of the maximum value of the angular frequency for simulations $\Omega_3$-b3.5e13 (blue), $\Omega_3$-b7e13 (red) and $\Omega_3$-b1e14 (green) with the \ac{MInIT} model. The grey overlapping lines depict the results from the same three simulations without including the \ac{MInIT} model. In all cases with the \ac{MInIT} model, we set $e_{\rm MRI}(0) = 10^{-10}e_{\rm kin}(0)$ and $e_{\rm PI}(0) = 10^{-11}e_{\rm kin}(0)$.}
    \label{fig:omg_max_tev}
\end{figure}

We compare in Figure~\ref{fig:eturb_tev_bcomp} the temporal evolution of the turbulent energies using different initial magnetic field strengths, $\bar{b}_0$, and keeping the initial angular frequency from the model $\Omega_3$ and the initial \ac{MRI} energy density amplitude to $e_{\rm MRI}(0)/e_{\rm kin}(0) = 10^{-10}$, and $e_{\rm PI}(0)/e_{\rm kin}(0) = 10^{-11}$. Although $e_{\rm MRI}$ evolves with the same rate, $e_{\rm PI}$  grows faster for smaller values of $\bar{b}_0$. This is explained by the dependence of the parasitic growth rate from Eq.~\eqref{grate_pi} with $\bar{b}_0$. The fact that $e_{\rm PI}$ grows faster leads to an earlier saturation of the energy densities at a lower amplitude for $\bar{b}_0 = 3.5\times 10^{13}$ G. This result is consistent with the findings made by~\cite{Obergaulinger-2008} and~\cite{Rembiasz:2016b}, who claimed that the \ac{MRI} amplification factor, defined as
\begin{equation}
    \mathcal{A} \equiv \frac{\sqrt{\bar{M}_{r\phi}(t_{\rm sat})}}{\bar{b}_0}\,,
\end{equation}
has a very weak dependence on the initial poloidal magnetic field. This translates in a linear dependence between $\sqrt{\bar{M}_{r\phi}(t_{\rm sat})}$ and $\bar{b}_0$, as depicted in Fig.~\ref{fig:esat_vs_bpol}. In this case, we show the maximum values of the square root of the averaged $r\phi$ component from the Maxwell stress tensor, over a radius of $r = 8$ km, for five different values of the initial poloidal magnetic field (see Table~\ref{table:omegas}). These maximum values clearly grow linearly with the poloidal field amplitude, $\bar{b}_0$. In Fig.~\ref{fig:omg_max_tev} we see the impact of the different choices of the initial magnetic field on the angular frequency of the \ac{NS}. It can be seen that, after the first $\sim 10$ ms, the maximum value of the equatorial angular frequency, $\Omega_{\rm max}$, starts decaying faster in those simulations with a larger poloidal magnetic field, once the \ac{MRI} reaches its largest amplitude. We note that the oscillations in the central rotation frequency at early times are due to the inclusion of the exponential decay of the density profile, which alters the equilibrium state of the \ac{NS}. Even though there is a slight decay of $\Omega_{\rm max}$ in the simulations without the \ac{MInIT} model (grey lines) due to numerical dissipation (see Appendix~\ref{app1}), this decay is much less pronounced than the rest of simulations which include the subgrid model. As mentioned above, since we have not included the subgrid terms in the induction equation, we do not expect any large-scale magnetic field amplification triggered by the \ac{MRI} dynamo. Therefore, the poloidal magnetic field, which is the component entering in Eqs.~\eqref{turb_en_mri} and~\eqref{turb_en_pi} through $\gamma_{\rm PI}$, remains nearly constant during the simulation.

\subsection{Dependence on the initial amplitudes of the turbulent energy densities}
\label{subsec::turb_amp}

We have the freedom to choose the initial value of the turbulent energy densities, $e_{\rm MRI}(0)$ and $e_{\rm PI}(0)$. The growth timescale and saturation amplitude of the \ac{MRI} might also be sensitive to the choice of these quantities. From the local analytical results from~\cite{Miravet:2024b} and the numerical results from~\cite{Rembiasz:2016b}, we should not expect a noticeable difference in the saturation amplitude of the instability. To test this, we employ simulation $\Omega_3$ from Table~\ref{table:omegas} with different choices of $e_{\rm MRI}(0)$ and $e_{\rm PI}(0)$, expressed as a function of the initial large-scale kinetic energy density, $e_{\rm kin}(0)$. 

\begin{figure*}
    \centering
    \includegraphics[width=\textwidth]{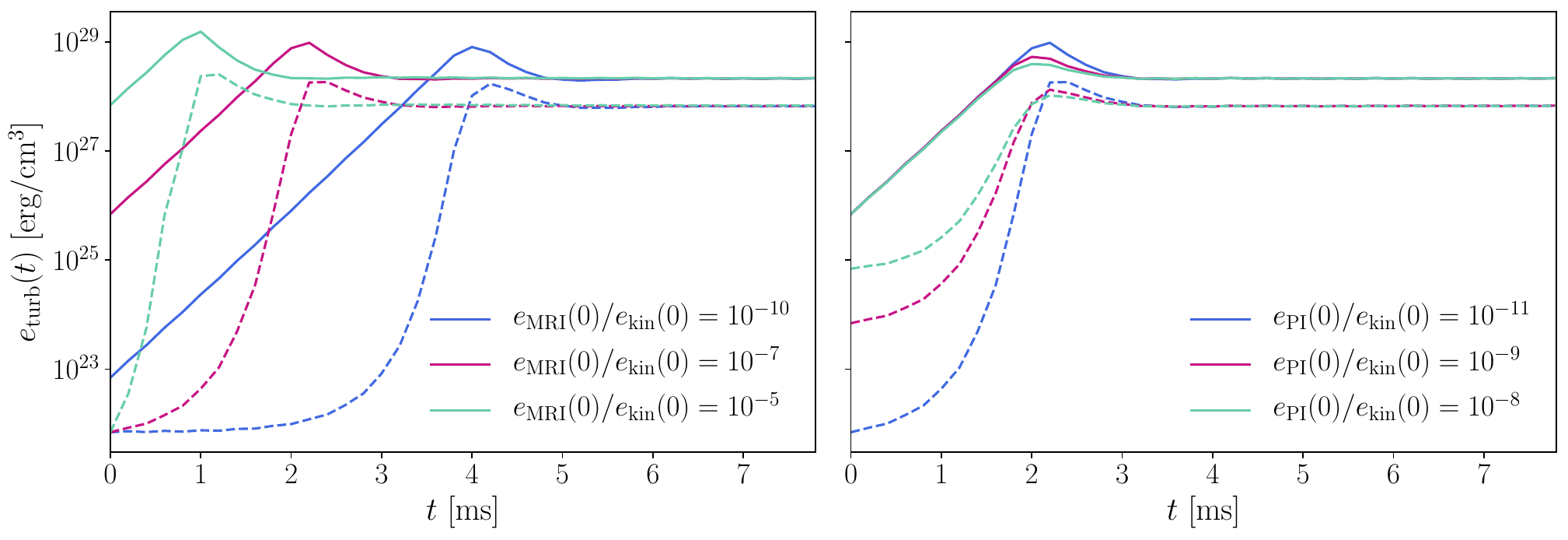}
    \caption{Time evolution of the turbulent energy densities, $e_{\rm MRI}$ (solid) and $e_{\rm PI}$ (dashed) averaged over a radius of $r = 8$ km. We show different choices for the initial turbulent energy densities, $e_{\rm MRI}(0)$ and $e_{\rm PI}(0)$. In the left panel we keep $e_{\rm PI}(0)$ fixed to $10^{-11}e_{\rm kin}(0)$, whereas in the right panel we set $e_{\rm MRI}(0) = 10^{-7}e_{\rm kin}(0)$. The central initial rotation frequency is fixed to $\Omega_3$.}
    \label{fig:eturb_tev_turbamps}
\end{figure*}

Figure~\ref{fig:eturb_tev_turbamps} depicts the time evolution of the turbulent energy densities for different initial amplitudes. The left panel shows different choices of $e_{\rm MRI}(0)$, fixing $e_{\rm PI}(0)$ at $10^{-11}e_{\rm kin}(0)$, whereas in the right panel we keep $e_{\rm MRI}(0)$ fixed at $10^{-7}e_{\rm kin}(0)$. We can observe in the left panel that choosing a different value for $e_{\rm MRI}(0)$ leads to a different saturation time. Larger values of the initial \ac{MRI} energy density lead to a more rapid growth of the parasitic energy density and therefore to an earlier saturation. Nevertheless, the saturation amplitude of the \ac{MRI} remains mostly insensitive to the choice of $e_{\rm MRI}(0)$. In the right panel of Fig.~\ref{fig:eturb_tev_turbamps} we see that simulations employing different values of $e_{\rm PI}(0)$ saturate at almost the same time. The maximum \ac{MRI} amplitude is slightly different~\citep[larger for lower amplitudes of the initial parasitic energy density, cf.][]{Miravet:2024b}, but after saturation all simulations reach the same \ac{MRI} and \ac{PI} values. These results lead to the conclusion that, assuming that the turbulent energy densities should be several orders of magnitude smaller than the large-scale kinetic energy density, the choice of their initial values should not have an important impact on the large-scale dynamics.

\section{Discussion}\label{sec::conclusions}

The lack of spatial resolution in current numerical simulations of \ac{BNS} mergers (and core-collapse supernovae) prevents the development of the \ac{MRI} during the post-merger phase. This undermines the credibility of the simulations as the \ac{MRI} can play a crucial role in the evolution of the remnant. Angular momentum transport drives the system toward rigid rotation, at which point the \ac{HMNS} is expected to collapse into a black hole. The characteristic timescale for angular momentum transport is estimated to be of $\mathcal{O}(100)$ ms, though it depends on the remnant’s rotational profile~\citep{Hotokezaka2013}. Hence, the \ac{BNS} remnant is expected to undergo collapse to a black hole within roughly $\mathcal{O}(100)$ ms after merger. This timescale has direct implications for the associated kilonova emission that powers $r$-process nucleosynthesis. Moreover, the collapse of the \ac{HMNS} is required for the launch of a \ac{GRB}, meaning that accurate simulations capable of reliably capturing the lifetime of the remnant are essential for interpreting gamma-ray detections from such mergers. Conversely, simulations that fail to incorporate turbulent effects may produce artificially long-lived remnants that cannot generate sufficiently powerful outflows to drive a \ac{GRB}.
  
The use of \ac{LES} can help capture the impact of the small-scale turbulence on the overall dynamics of the system. In this work we have presented results from simulations of differentially rotating, magnetised \acp{NS} including the \ac{MInIT} subgrid model we first presented in~\citet{Miravet:2022}. With the addition of the evolution equations of the turbulent energy densities for the \ac{MRI} and the \ac{PI} from the \ac{MInIT} model, we have been able to compute the turbulent stress tensors responsible for turbulent angular momentum transport, i.e.,~the Maxwell and Reynolds stresses. We have observed that their inclusion in the momentum equation of the fluid has an impact on the global dynamics of the \ac{NS}, leading to a net transport of angular momentum radially outwards, which reduces the rotation frequency of the star in its central regions. Moreover, in the regions where the rotational profile flattens, the turbulent energy densities and, therefore, the turbulent stresses, decay, as one would expect when the fluid loses its differential rotation. 

Our results show that the evolution of the turbulent energy densities is sensitive to the rotation frequency of the star and to the strength of the poloidal magnetic field. Although the saturation amplitude of the energies is very similar, different choices of the central rotation frequency of the \ac{NS} lead to different growth rates of the \ac{MRI}, which results in different saturation times. Regarding the amplitude of the initial magnetic field, its choice has a direct impact on the growth of the \ac{PI}, resulting in a larger parasitic growth rate for lower magnetic field amplitudes. Thus, stronger poloidal fields lead to a larger saturation amplitude, as previously noted by~\cite{Rembiasz:2016a,Rembiasz:2016b} and~\cite{Miravet:2024b}, which in turns produces a more effective angular momentum transport.

The \ac{MInIT} model also needs a value of the initial amplitudes of the turbulent energies. Assuming that these quantities will be a small fraction of the initial large-scale kinetic energy, we have performed several simulations with different choices of these initial values. Although they impact the saturation time of the instability, the saturation amplitude remains the same. Therefore, the effect of the choice of the initial amplitude of the turbulent energies on the growth of the \ac{MRI} is somewhat similar to that produced by different values of the central rotation frequency of the \ac{NS}. One might think this could be an issue, since $e_{\rm MRI}(0)$ and $e_{\rm PI}(0)$ are free parameters of the model. Nevertheless, their impact is fairly low as this just delays saturation by a few milliseconds. 

It is important to point out that this work should be regarded as a promising test of the \ac{MInIT} model in global simulations of isolated \acp{NS}. However, there are a number of assumptions that should be relaxed in future work. On the one hand, the setup of the simulations is far from being realistic: we enforced axisymmetry, employed Newtonian dynamics, used a polytropic \ac{EOS}, and adopted a rotation law which, although widely employed in the literature, departs from those inferred from simulations for \ac{BNS} merger remnants~\citep{Hanauske:2017,Iosif:2022,Cassing:2024}. Those are all aspects that need to be considered in future applications of the model. Furthermore, the most immediate step to make to improve the model is the implementation of the Faraday stress tensor in the induction equation for the magnetic field, which might induce an effective dynamo. This could also have an impact on the evolution of the \ac{MRI} itself after saturation. 
In addition, it would be advisable to conduct a comparison between low-resolution simulations including the \ac{MInIT} model and high-resolution simulations able to capture the \ac{MRI} unaided by a subgrid model, and in three dimensions. In this regard we note that since our simulations are axisymmetric, the associated \ac{MRI}  would be quantitatively different from the one expected in 3D~\citep{Obergaulinger:2009}, which is the one used to calibrate the \ac{MInIT} model in~\cite{Miravet:2022}. However, in order to run a 3D simulation that fully resolves the \ac{MRI}, we would need  enough spatial resolution to cover at least 10 grid cells per \ac{MRI} wavelength. As an illustrative example, the simulation $\Omega_3$, with $\bar{b}_0 = 7\times 10^{13}$ G, has a radial resolution of $107$ m, whereas $\lambda_{\rm MRI}$ is found to be $\sim 10$ m through almost the whole stellar interior. The total CPU time of the simulation was $\approx 300$ hours. To fully resolve the \ac{MRI} in 3D, we would need to 
add the azimuthal dimension and increase the resolution by a factor 100 in each direction, which would increase the CPU time by a factor $\sim 10^8$, requiring $\sim10^6$ years of CPU time.

Finally, the use of Newtonian physics and dynamics is a major limitation of our current approach.
In \ac{BNS} mergers (and also core-collapse supernovae) the spacetime metric deviates strongly from the flat metric, and the fluid velocity can become relativistic in some regions. Therefore, a more realistic approach would be the performance of fully general-relativistic \ac{LES}, as done in, e.g.,~\citet{Giacomazzo:2015},~\citet{Radice:2020},~\citet{Palenzuela:2022} and~\citet{Aguilera-Miret:2025}. Nevertheless, most of these works solely focus on the turbulent magnetic field amplification by the \ac{KHI} and are unable to capture the effects of the \ac{MRI} in the post-merger phase. Moreover, they lack the ability to handle the dependence of the saturation amplitude on the magnetic field amplitude, the relation between the rotation frequency and the growth phase, or the decay of turbulence in MRI-stable regions. Another key issue is that subgrid models applied to \ac{GRMHD} simulations of \ac{BNS} mergers are covariant with respect to transformations in the spatial coordinates, but not when it comes to general spacetime coordinate transformations~\citep{Duez:2020,Radice:2024}. Non-covariant closures can introduce coordinate-independent artifacts in the simulations, since the averaging applied to a single foliation can inherit the dependencies of that spacetime
slice. For that purpose, it is important to develop a covariant approach~\citep{Duez:2020,Celora:2021,Celora:2024a,Celora:2024b} that would allow the \ac{MInIT} model to be used in general-relativistic simulations.  We plan to report on those extensions of the model in future work.

\section*{Acknowledgements}

This work has been supported by the Spanish Agencia Estatal de Investigación (grants PID2021-125485NB-C21 and PID2024-159689NB-C21) funded by MCIN/AEI/10.13039/501100011033 and ERDF A way of making Europe, and by the Prometeo excellence programme grants CIPROM/2022/13 and CIPROM/2022/49 funded by
the Generalitat Valenciana. MMT acknowledges support from the Science and Technology Facilities Council (STFC), via grant No.~ST/Y000811/1, and from the Ministerio de Ciencia, Innovación y Universidades del Gobierno de España through the ``Ayuda para la Formación de Profesorado Universitario'' (FPU) fellowship No.~FPU19/01750. MO was supported by the Ram\'on y Cajal programme of the Agencia Estatal de Investigaci\'on (RYC2018-024938-I).
 MR acknowledges support by the Generalitat Valenciana grant CIDEGENT/2021/046. We acknowledge further
support from the European Horizon Europe staff ex-
change (SE) programme HORIZON-MSCA2021-SE-01
Grant No. NewFunFiCO-101086251.
 
\section*{Data Availability}
 
The data underlying this article will be shared on reasonable request to the corresponding author. 


\bibliographystyle{mnras}
\bibliography{biblio}



\appendix

\section{Numerical dissipation}
\label{app1}

The discretisation of partial differential equations for numerical solutions inevitably introduces numerical dissipation, which affects the evolution of the system. This phenomenon is well understood and, among other consequences, leads to a reduction in kinetic energy. As illustrated in Fig.~\ref{fig:omg_max_tev}, numerical dissipation may also explain the reduction in the maximum rotation frequency observed in MHD simulations that exclude MInIT (grey lines).

\begin{figure}
    \centering
    \includegraphics[width=\linewidth]{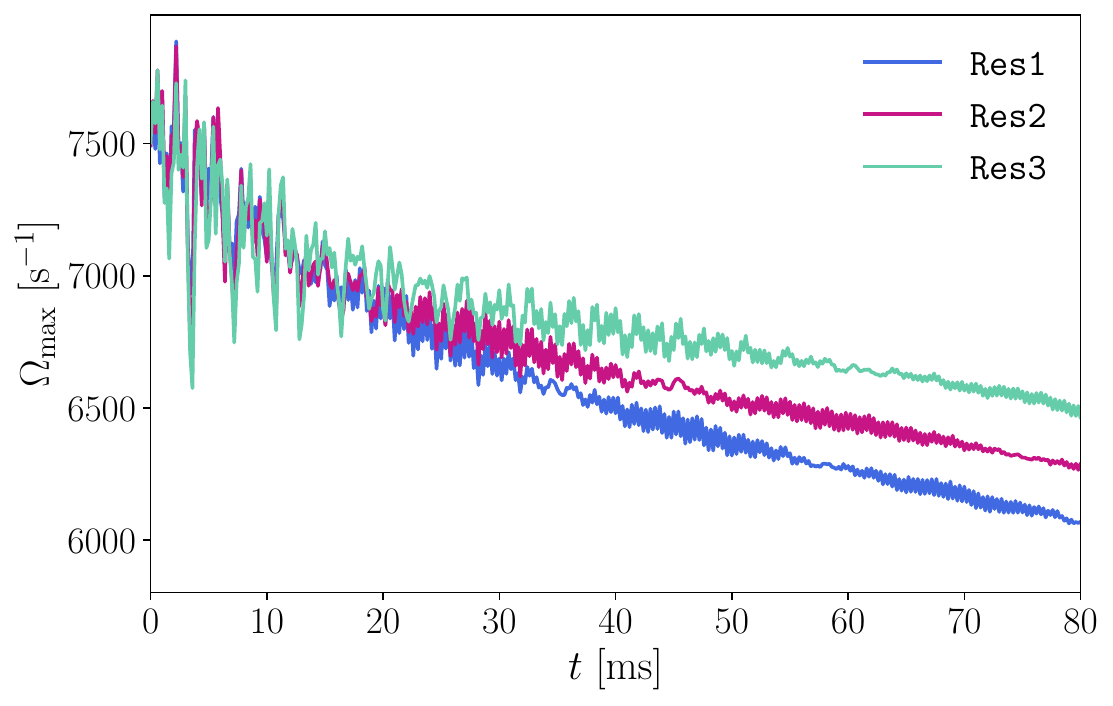}
    \caption{Time evolution of the maximum value of the angular frequency, $\Omega_{\rm max}$, at the equator for the simulation $\Omega_3$, with three different spatial resolutions. The simulation with the largest resolution, \texttt{Res3}, shows a slower decay of $\Omega_{\rm max}$ compared to the simulations with a lower resolution.}
    \label{fig:app_omg}
\end{figure}

To test this hypothesis, we performed two additional simulations with finer spatial resolutions: \texttt{Res2}, with a grid $(N_r,N_{\theta}, N_{\phi}) = (576,252,1)$, and \texttt{Res3}, with $(720,360,1)$. For comparison, we also include the baseline simulation already used in this work, $\Omega_3$ (labelled here as \texttt{Res1}),with $(N_r,N_{\theta}, N_{\phi}) = (468,180,1)$. Figure~\ref{fig:app_omg} demonstrates that higher spatial resolution reduces the decline of the maximum equatorial angular frequency over time. However, a decay remains, indicating that kinetic energy is still dissipated through grid discretisation. The time evolution of $\Omega_{\rm max}$ is qualitatively consistent across all runs, with maximum relative differences of $\sim 3\%$ between \texttt{Res2} and \texttt{Res3}, and $\sim 5\%$ between \texttt{Res1} and \texttt{Res3}.

A fully convergent simulation with negligible numerical dissipation would show no decrease in $\Omega_{\rm max}$. In practise, however, the limited resolution and the out-of-equilibrium initial conditions (arising from modified density profiles) prevent complete convergence. Still, the small relative differences across resolutions give us confidence that the results from \texttt{Res1} remain qualitatively robust.

\begin{figure}
    \centering
    \includegraphics[width=\linewidth]{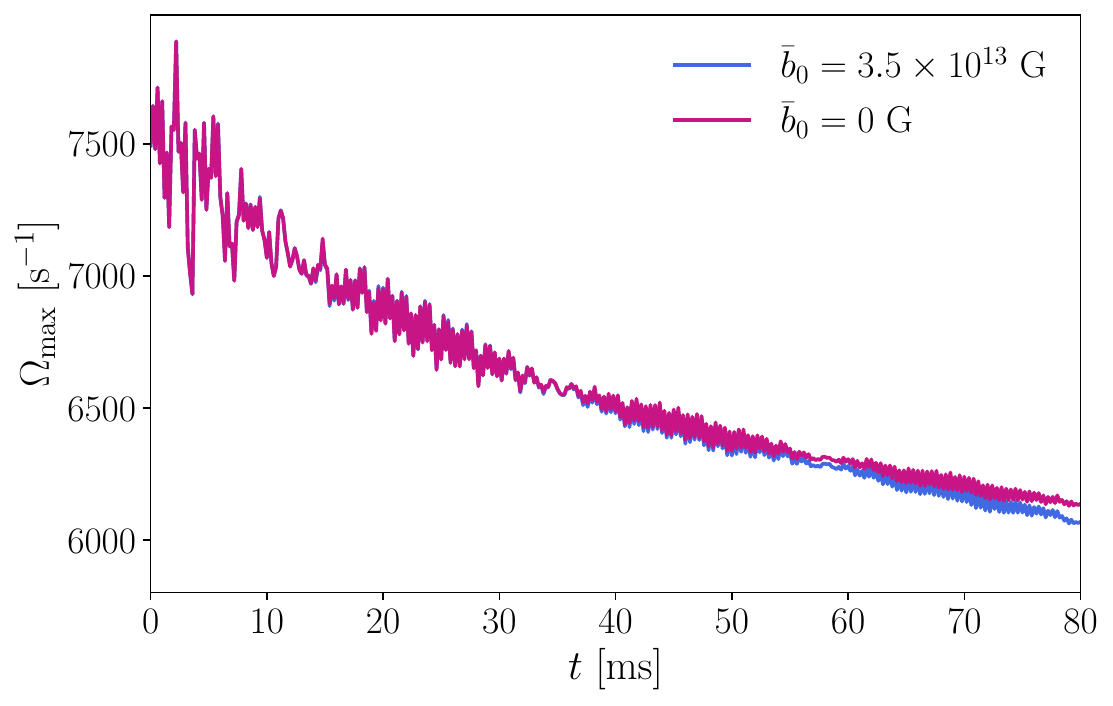}
    \caption{Time evolution of the maximum value of the angular frequency, $\Omega_{\rm max}$, at the equator for the simulation $\Omega_3$, with (blue) and without magnetic fields (red). The evolution is identical in both cases, except for late times ($t>$ 60 ms), where the higher numerical viscosity of the magnetised case leads to a larger decay of $\Omega_{\rm max}$.}
    \label{fig:app_bfield}
\end{figure}

One could argue that the decrease in $\Omega_{\rm max}$ might have a physical origin, perhaps due to the action of a partially resolved MRI. This is excluded by construction: the chosen spatial resolution ensures that the cell size is roughly ten times larger than the wavelength of the fastest-growing MRI mode, $\lambda_{\rm MRI} = 2\pi / k_{\rm MRI}$ (see Eq.~\eqref{wavenumber}). We further confirm this by performing a simulation without magnetic fields. In Fig.~\ref{fig:app_bfield}, we show the equatorial time evolution of $\Omega_{\rm max}$ for simulation $\Omega_3$ from Table 1, both with (red) and without (green) magnetic fields. The two curves are indistinguishable up to $t \approx 60$ ms, after which they diverge slightly. This minor discrepancy arises because numerical viscosity is generally higher in \ac{MHD} simulations than in purely hydrodynamical ones.

\bsp	
\label{lastpage}
\end{document}